\documentclass[letterpaper,11pt]{article}
\usepackage{amsthm}  
\usepackage{amsmath, amssymb}
\usepackage{color}
\usepackage{fullpage}
\usepackage[numbers]{natbib}
\usepackage[noend]{algorithmic}
\usepackage{thmtools, thm-restate}
\usepackage{tikz}
\usepackage{enumerate}
\usepackage{thm-restate}
\usepackage{graphicx}
\usepackage{amsopn}
\usepackage{caption}
\usepackage{hyperref}
\usepackage{subcaption}
\usepackage{zerosum,csquotes}
\usepackage{enumitem,linegoal}
\usepackage[linesnumbered,ruled,vlined]{algorithm2e}
\usepackage{comment}
\usepackage{float}

\usepackage{ctable}					

\usepackage{mathtools}
\usepackage[flushleft]{threeparttable}
\usepackage[normalem]{ulem}

\usepackage{footnote}
\makesavenoteenv{tabular}
\makesavenoteenv{table}

\usepackage[margin=1in]{geometry}

\makeatletter
\newif\ifqed
\def\GrabProofArgument[#1]{ #1: \egroup\ignorespaces}
\def\proof{\noindent\textbf\bgroup Proof%
	\@ifnextchar[{\GrabProofArgument}{. \egroup\ignorespaces}\global\qedtrue}

\def\qedhere{\ifmmode\tag*{\qedsign}\else\hspace*{\fill}\qedsign\medskip\fi\global\qedfalse}
\def\qedsign{$\Box$}
\makeatother

\usepackage{float}
\usepackage{enumitem,linegoal}
\usepackage{tabularx}
\usepackage{booktabs}
\newcolumntype{H}{>{\setbox0=\hbox\bgroup}c<{\egroup}@{}}

\usepackage{thmtools}
\usepackage{thm-restate}
\usepackage[normalem]{ulem}
\usepackage{filecontents}
\usepackage{float}
\usepackage{tabularx}
\usepackage{booktabs}
\newcolumntype{H}{>{\setbox0=\hbox\bgroup}c<{\egroup}@{}}

\usepackage[draft]{fixme}
\fxsetup{mode=multiuser,theme=color, layout=inline}
\FXRegisterAuthor{ss}{ass}{\color{red} \bf Saeed}
\FXRegisterAuthor{tk}{atk}{\color{blue} \bf  Tomasz}

\newtheorem{theorem}{Theorem}

\newtheorem{corollary}[theorem]{Corollary}
\newtheorem{definition}{Definition}

\newcounter{theo}[section] 

\definecolor{mygreen}{RGB}{20,140,80}
\definecolor{mylightgray}{RGB}{230,230,230}

\definecolor{mygreen}{RGB}{20,140,80}
\definecolor{mydarkgray}{gray}{0.15} 
\definecolor{oceanblue}{HTML}{2c55c2}
\hypersetup{
     colorlinks=true,
     citecolor= mygreen,
     linkcolor= black
}

\SetCommentSty{mycommfont}

\newcommand*\samethanks[1][\value{footnote}]{\footnotemark[#1]}

\newcounter{proccnt}

\newcommand{\konote}[1]{}
\definecolor{printable}{RGB}{43,70,120}

\title{Optimal Space and Time for Streaming Pattern Matching}
\author{
	Tung Mai\thanks{Adobe Research. Email: \texttt{\{tumai,anuprao,rrossi\}@adobe.com}.}
	\and
	Anup Rao\samethanks[1]
	\and
	Ryan A. Rossi\samethanks[1]
	\and
	Saeed Seddighin\thanks{Toyota Technological Institute. Email: \texttt{saeedreza.seddighin@gmail.com}}
}

\begin{document}
\renewcommand{\theenumi}{(\roman{enumi})}
\renewcommand{\labelenumi}{\theenumi.}
\sloppy

\date{}

\maketitle

\begin{abstract}
In this work, we study longest common substring, pattern matching, and wildcard pattern matching in the asymmetric streaming model. In this streaming model, we have random access to one string and streaming access to the other one. We present streaming algorithms with provable guarantees for these three fundamental problems. In particular, our algorithms for pattern matching improve the upper bound and beat the unconditional lower bounds on the memory of randomized and deterministic streaming algorithms. In addition to this, we present algorithms for wildcard pattern matching in the asymmetric streaming model that have optimal space and time.
\end{abstract}

\section{Introduction}
We consider problems of pattern matching, wildcard pattern matching, and  longest common substring  in the asymmetric streaming model. Pattern matching, also sometimes referred to as substring search, is a fundamental and widely used string operation. In this problem, a text $T$ and a pattern $P$ are given as input and the goal is to search for occurrences of the pattern in the text. We also study two classical generalizations of this problem. Pattern matching with wild card is  when we allow '?' character in the pattern that can match with any character in the text.  Longest common substring is another classic problem in which one is given two strings and the task is to find the longest consecutive sequence of characters that appear in both of the strings. We study these fundamental string problems in the recently introduced  asymmetric streaming model. In this setting, we have random access to one of the strings and streaming access to the other one. We give near optimal algorithms for all these problems. 

Popular undergraduate textbooks on algorithms \cite{cormen2009introduction, sedgewick2002algorithms} have chapters with various algorithms for the pattern matching problem.  \cite{knuth1977fast, karp1987efficient} are two well-known and classic papers that give linear time algorithms in the case when both the text and the pattern strings are available offline. The study of longest common substring led to the discovery of some of the most important data structures in algorithms, namely \textit{suffix arrays} and \textit{suffix trees}. \cite{cole2002verifying} gives the current best algorithm for pattern matching with wildcard in the offline setting.

These algorithms are not suitable for the case when the strings are too large to store in the memory. Streaming is a popular computational model used to study algorithms with low memory constraints. In the streaming setting, input data arrives as a data stream and there is only sublinear memory available to the algorithm~\cite{liben2006finding,SW07,belazzougui2016edit,CGK16}.
Streaming algorithms are designed to solve exact or approximate the solution using only a few passes (rounds) over the data.
There have been some works that have focused on related problems in streaming data, including longest common subsequence~\cite{liben2006finding,DBLP:conf/focs/GalG07}, edit distance~\cite{belazzougui2016edit}, among others~\cite{DBLP:conf/soda/GopalanJKK07,naumovitz2014polylogarithmic,DBLP:conf/soda/ErgunJ08}.

The asymmetric streaming model~\cite{andoni2010polylogarithmic,saks2013space} was introduced by Andoni et al.~\cite{andoni2010polylogarithmic} and Sakes and Seshadhri~\cite{saks2013space}.
In this streaming model, we have random access to one string and streaming access to the other. Similar to the streaming setting, the goal here is to design an algorithm that reads the streaming part in a few (constant) rounds and uses a small memory to solve/approximate the problem. In this work, we study our problems in this new setting and present algorithms/lower bounds for each of the problems.

\begin{table*}[h!]
\centering
\small
\begin{tabular}{@{}l @{} c@{}ccccH@{}}
\midrule
\textbf{Problem} & 
\textbf{Solution} & 
\textbf{Rounds} &
\textbf{Runtime} & 
\textbf{Space} & 
\textbf{Reference} 
\\
\midrule

\text{Asym. Streaming Pattern Matching} & 
Exact & 
$1$ &
$O(n+m)$ & 
$O(1)$ & 
Theorem~\ref{thm:asym-stream-pattern-matching} \\

\text{Asym. Streaming LCS}
& Exact & 
$1$ &
$\tilde O(nm)$ & 
$O(1)$ & 
Theorem~\ref{thm:asym-stream-longest-common-substring} \\

& $1-\epsilon$ & 
$O(1/\kappa)$ &
$\tilde O(n^{2+\kappa}/d)$ & 
$O(\log n / \epsilon)$ & 
Theorem~\ref{theorem:lcs2} 
\\

& $1-\epsilon$ & 
$O(\log n / \epsilon)$ &
$\tilde O(n^2/d)$ & 
$O(1 / \epsilon)$ & 
Corollary~\ref{cor:lcs3} \\

\text{Asym. Streaming Wildcard Pattern Matching} & 
Exact & 
$1$ &
$-$&
$\Omega(n)$ & 
Theorem~\ref{thm:asym-streaming-wildcard-pattern-matching-memory} \\

\text{Random Access Wildcard Pattern Matching} & 
Exact & 
$-$ &
$\tilde{O}(n^2/s)$&
$O(s)$ & 
Theorem~\ref{thm:non-streaming-wildcard-pattern-matching} \\

\midrule
\end{tabular}
\caption{Summary of our key results and algorithms. 
}
\label{table:summary-key-results}
\vspace{-2mm}
\end{table*}

\subsection{Preliminaries}
Here we formally define the problems that we discuss in this manuscript. In the pattern matching problem, a text $T$ is given via oracle queries and a pattern $P$ comes as a stream of characters. The goal is to verify if $P$ appear in $T$ as a substring. We denote by $n=|T|$ and $m=|P|$, the length of the text and pattern strings. 

\begin{definition}[Asymmetric Pattern Matching]
	Let $T$ be a text available via value queries. That is in each query we can give an index $i$ to an oracle and in return we earn the $i$'th character of the text. Also, a pattern $P$ comes as a stream of characters. The goal of the pattern matching problem is to find out if pattern $P$ appears as a substring of string $T$ or not.
\end{definition}

In the wildcard variant of the problem, the pattern is allowed to also include a special character '?' which may appear in several places. 
These characters can match with any character of the text.
A slightly generalized version of pattern matching is \textit{the longest common substring} or in short the \textsf{LCS} problem. In \textsf{LCS}, two strings $A$ and $B$ are provided and the goal is to find the largest substring of $A$ which also appears in $B$. Unlike pattern matching, \textsf{LCS} is symmetric; If we exchange $A$ and $B$, the answer remains the same. 
In the asymmetric streaming variant of the problem, we assume one string is available via oracle queries and the other one comes as a stream of characters.

\begin{definition}[Asymmetric Longest Common Substring]
	Let $A$ be a string available via value queries. 
	That is in each query we can give an index $i$ to an oracle and in return learn the $i$'th character of $A$. 
	Also, another string $B$ comes as a stream of characters. 
	In the \textsf{LCS} problem, we are interested in the largest substring shared between $A$ and $B$.
\end{definition}

\subsection{Key Results}
In this paper, we give a randomized $O(n + m)$-time and $O(1)$ space algorithm for asymmetric streaming pattern matching. Our algorithm is optimal both in terms of runtime and memory.

\vspace{0.2cm}
{\noindent \textbf{Theorem~\ref{thm:asym-stream-pattern-matching}} (restated informally). 
\textit{There exists a randomized algorithm for asymmetric streaming pattern matching that solves the problem with memory $O(1)$ and runtime $O(n + m)$. The algorithm runs in a single pass and succeeds with probability at least $0.99$.\\}}
\vspace{-0.4cm}

Since the algorithm of Theorem~\ref{thm:asym-stream-pattern-matching} is randomized, we also present a deterministic variant of our algorithm that runs with $O(1)$ memory. However, the runtime of this algorithm grows to quadratic. Our result is in contrast to the lower bound of $\Omega(m)$ on the memory of deterministic streaming algorithms for pattern matching.

\vspace{0.2cm}
{\noindent \textbf{Corollary~\ref{cor:cor}} (restated informally). 
\textit{There exists a deterministic algorithm for asymmetric streaming pattern matching that solves the problem with memory $O(1)$ and runtime $O(nm)$.\\}}
\vspace{-0.4cm}

For pattern matching with wild card, we give an impossibility result. We show  that there is no exact $o(n)$ memory algorithm for asymmetric streaming pattern matching with wildcard. We therefore consider the random access model, where we have random access to characters of both the text and the pattern. In this model, we give an algorithm that requires sublinear memory. We give an algorithm for wildcard pattern matching in the non-streaming setting that uses $O(s)$ space and takes $\tilde O (n^2/s)$ time, where $s$ can vary from $1$ to $n$. These results are summarized in Table~\ref{table:summary-key-results}. 

\vspace{0.2cm}
{\noindent \textbf{Theorem~\ref{thm:asym-streaming-wildcard-pattern-matching-memory}} (restated informally). 
\textit{Any one pass asymmetric streaming algorithm for wildcard pattern matching requires memory $\Omega(n)$.\\}}
\vspace{-0.4cm}

\vspace{0.2cm}
{\noindent \textbf{Theorem~\ref{thm:non-streaming-wildcard-pattern-matching}} (restated informally). 
\textit{For any $1 \leq s \leq n$ there exists an $\tilde O(n^2/s)$ algorithm for wildcard pattern matching with space $O(s)$ in the non-streaming setting. \\}}
\vspace{-0.4cm}

We also give asymmetric streaming algorithms for longest common substring. This algorithm also requires $O(1)$ memory and runs in quadratic time. Since the runtime of \textsf{LCS} in the sequential setting is linear, we seek to tighten the gap between the two runtimes. We prove that when the solution size $d$ is large, we can approximate the solution with low memory and in subquadratic time.

\vspace{0.2cm}
{\noindent \textbf{Theorem~\ref{thm:asym-stream-longest-common-substring}} (restated informally). 
\textit{There exists a randomized algorithm for asymmetric streaming \textsf{LCS} that solves the problem with memory $O(1)$ and runtime $\tilde O(nm)$. This algorithm succeeds with probability at least $1-n^{-5}$.\\}}
\vspace{-0.4cm}

\vspace{0.2cm}
{\noindent \textbf{Theorem~\ref{theorem:lcs2}} (restated informally). 
\textit{For any $\epsilon, \kappa > 0$, there exists an algorithm for \textsf{LCS} that approximate the solution within a factor $1-\epsilon$ in $O(1/\kappa)$ rounds with memory $O(\log n/\epsilon)$ and its overall runtime is bounded by $\tilde O(n^{2+\kappa}/d)$ in the worst case.\\}}
\vspace{-0.4cm}

\vspace{0.2cm}
{\noindent \textbf{Corollary~\ref{cor:lcs3}} (restated informally). 
\textit{For any $\epsilon> 0$, there exists an asymmetric streaming algorithm for \textsf{LCS} that approximate the solution of \textsf{LCS} within a factor $1-\epsilon$ in $O(\log n/\epsilon)$ rounds with memory $O(1/\epsilon)$  and its overall runtime is bounded by $\tilde O(n^2/d)$ in the worst case.\\}}

\subsection{Related Work} \label{sec:related-work}
In the streaming setting, input data arrives as a data stream and there is only sublinear memory available to the algorithm~\cite{liben2006finding,SW07,belazzougui2016edit,CGK16}.
This setting has become increasingly important for modeling memory (space) constraints.
Streaming algorithms are designed to solve exact or approximate the solution using only a few passes (rounds) over the data.
There have been some works that have focused on related problems in streaming data, including longest common subsequence~\cite{liben2006finding,DBLP:conf/focs/GalG07}, edit distance~\cite{belazzougui2016edit}, among others~\cite{DBLP:conf/soda/GopalanJKK07,naumovitz2014polylogarithmic,DBLP:conf/soda/ErgunJ08}.

The asymmetric streaming model~\cite{andoni2010polylogarithmic,saks2013space} was introduced by Andoni et al.~\cite{andoni2010polylogarithmic} and Sakes and Seshadhri~\cite{saks2013space}.
In this streaming model, we have random access to one string and streaming access to the other.
Recently, Farhadi et al.~\cite{farhadi2020streaming} presented asymmetric streaming algorithms for edit distance and longest common subsequence.
Notably, that work presents an algorithm with a constant factor approximation for edit distance with memory $\tilde O(n^{\delta})$ for any $\delta>0$.
In contrast, our work focuses on asymmetric streaming algorithms for pattern matching, wildcard pattern matching, and longest common substring.

In terms of the streaming pattern matching problem, 
\cite{clifford2011black} proves that any deterministic streaming algorithm requires memory $\Omega(n)$ whereas
\cite{porat2009exact} gives a randomized streaming algorithm that requires only memory $O(\log m)$. 
Recent work has focused on approximate streaming methods for matching multiple patterns in multiple streams~\cite{golan2018towards,golan2017real}.
For wildcard pattern matching in the streaming model, 
\cite{golan2019streaming,golan2017real} focuses on pattern matching with d wildcards whereas
\cite{ergun2020periodicity} studies periodicity in streaming model with wildcards.

\section{Pattern Matching}
In this section we discuss an asymmetric streaming algorithm for pattern matching. This algorithm is optimal in the sense that it only uses $O(1)$ memory and its time complexity is linear ($O(|T|+|P|)$). Our algorithm is randomized and succeeds with high probability. That is, if the pattern is not appeared in the text, the output of our algorithm is always negative. Also, if the pattern does appear in the text, then with probability $0.99$ our algorithm gives a positive output. The algorithm is described below:

Let $|T|^2 < q \leq 2|T|^2$ be a random prime number and $f: \Sigma \rightarrow [0,|\Sigma|-1]$ be any arbitrary function that maps the characters of the alphabet to distinct consecutive numbers starting from $0$. We define the value realization of a string $S$ (a.k.a its hash value) as the number we obtain by replacing the characters by their corresponding values in $f$ taking the overall value in base $|\Sigma|$. More precisely, the value realization of a string $S$ is equal to $$ \sum |\Sigma|^{i-1} f(S[i]) .$$ We refer to this expression as $f(S)$.

In our algorithm, at each point in time we have to pointers $\ell$ and $r$ and a value $r$. $\ell$ and $r$ represent represent the \textbf{leftmost} appearance of the \textbf{current} pattern in $T$ and $r$ represents its value realization module $q$. The main reason we only keep the module $q$ value is that this can be quite large and we may need memory $\omega(1)$ to store the whole hash value. If at some point we conclude that $P$ does not appear in $T$, we terminate the algorithm.

Upon the arrival of each element, we update the hash value of the pattern in time $O(1)$ and verify whether it matches with $T[\ell, r+1]$. If so, no further action is required. Otherwise, we move both $\ell$ and $r$ to the right and again verify if the hash value module $q$ is equal to $r$. It follows from the mathematical formulation of the hash values that when we move the pointers by one, we can update the hash values in time $O(1)$. In our algorithm, we continue on by moving the pointers to the right until we find the pattern in the text. If we reach the end of the text before finding the pattern we conclude that the pattern does not appear in the text at all.

\begin{algorithm}[h!]
	\SetAlgoLined
	\KwResult{}
	online-hash $\leftarrow 0$\;
	offline-hash $\leftarrow 0$\;
	$\ell \leftarrow 1$\;
	$r \leftarrow 0$\;
	\While{new character comes}{
		update online-hash\;
		$r \leftarrow r+1$\;
		update offline-hash\;
		\While{online-hash $\neq$ offline-hash}{
			$\ell \leftarrow \ell +1$\;
			$r \leftarrow r+1$\;
			\If {$r > $ length of the offline string}{
				Report no match exists between the pattern and the text\;
			}
			update  offline-hash\;
		}
	}
    Report interval $[\ell, r]$\;
	\caption{Exact asymmetric streaming algorithm for pattern matching}\label{alg:pattern-matching}
\end{algorithm}

\begin{theorem}\label{thm:asym-stream-pattern-matching}
	There exists a randomized algorithm for asymmetric streaming pattern matching that solves the problem with memory $O(1)$ and runtime $O(n+m)$. The algorithm succeeds with probability at least $0.99$.
\end{theorem} 
\begin{proof}
	We outlined the algorithm above. Here we discuss its correctness and its complexity. We assume throughout this proof that when the hash values of two strings module $q$ are the same then the two strings are also the same. We explain later that this incurs an error only with a small probability.
	
	The guarantee of our algorithm is that $T[\ell,r]$ is the leftmost appearance of $P$ in $T$. Thus, whenever $P$ gets updated, we only need to take into account the intervals whose starting positions are not before $\ell$. We move the pointers to the right one by one and test if the new interval matches with $P$. Thus, this way we find the leftmost appearance of the updated pattern.
	
	The time complexity is bounded by $O(|T|+|P|)$. Notice that we only move the pointers in one direction and each move takes time $O(1)$. Thus, the overall time complexity for the moves is bounded by $O(|T|)$. Also, each time a new character is added to $P$, we update the hash value in time $O(1)$ so this gives an additional additive $O(|P|)$ to the runtime. Therefore, the overall time complexity is $O(|T|+|P|)$. Also, it follows from the algorithm that the memory complexity remains $O(1)$ throughout the process.
	
	Finally, we address the error of the hashing technique. Since $q$ is chosen randomly, the probability that two unequal strings end of with the same hash value module $q$ is roughly $\Theta(1/q)$. Therefore, by choosing $q$ in range $[|T| ^2, 2|T|^2]$ we can be sure that our assumption is correct with probability $\Theta(1- 1/|T|^2)$ in every step and therefore by union bound the total error probability is bounded by  $\Theta(1- 1/|T|)$.
\end{proof}

We conclude this section by bringing one more observation. The algorithm of Theorem~\ref{thm:asym-stream-pattern-matching} is randomized. However, one can de-randomize the algorithm via multiplying the runtime by a factor $O(|P|)$. That is, every time the hash-values are the same, we spend an additional $O(|T|)$ time to verify whether the two substrings are really equal. This makes the algorithm deterministic while it does not change the memory of the algorithm. This is in contrast with the streaming setting since no deterministic algorithm can solve pattern matching with memory $o(|P|)$.

\begin{corollary}[of Theorem~\ref{thm:asym-stream-pattern-matching}]\label{cor:cor}
	There exists a deterministic algorithm for asymmetric streaming pattern matching that solves the problem with memory $O(1)$ and runtime $O(nm)$.
\end{corollary}

\section{Wildcard Pattern Matching}
Motivated by our optimal solution for asymmetric streaming pattern matching problem, we consider a generalization of the pattern matching problem namely \textit{wildcard pattern matching}. We show in this section that unlike pattern matching, any asymmetric streaming algorithm for the wild card setting requires memory $\Omega(n)$. We accompany this result by improved algorithms for wildcard pattern matching for restricted settings.
The main result of this section is summarized in the following:

\begin{theorem} \label{thm:asym-streaming-wildcard-pattern-matching-memory}
	Any one pass asymmetric streaming algorithm for wildcard pattern matching requires memory $\Omega(n)$.
\end{theorem}
\begin{proof}
Consider the following text
	\[ \underbrace{11\ldots1}_k0\underbrace{11\ldots1}_k0\]
	and pattern
	\[  \underbrace{aa \ldots a}_k \underbrace{11\ldots1}_l0.  \]
	Here $a$ is either 1 or $?$ and $l$ is an integer between 1 and $k$. 
	We will show that the first $k$ characters of the pattern need to be stored exactly to give a correct answer.  
	Suppose the $i$-th character in the pattern, for some $i \leq k$, is not known to be 1 or $?$.
	Setting $l=i$ gives a pattern that matches the text if and only if the $i$-th character is $?$.  
	Therefore, any one pass asymmetric streaming algorithm for wildcard pattern matching must store the first $k$ positions of the pattern in this example. The theorem follows since $k$ is linear in the size of the input. 
\end{proof}

\section{Non-Streaming Wildcard Pattern Matching}

The previous section states an impossibility result for the wildcard pattern matching problem. In particular, memory $\Omega(n)$ is required in the asymmetric streaming setting. Therefore, to obtain low-space algorithms, we consider the non-streaming setting, where both pattern and text are fixed, and we have random access to every character of them. 
In this setting, a simple algorithm that considers all possible positions takes $O(1)$ space and $O(n^2)$ time. On the other hand, algorithms based on convolution \cite{Fischer1974STRINGMATCHINGAO, KalaiDontCares, CLIFFORD200753, Indyk98} take $O(n)$ space and $\tilde O(n)$ time. We give an algorithm (Algorithm 2) for wildcard pattern matching in the non-streaming setting that uses $O(s)$ space and takes $\tilde O (n^2/s)$ time, where $s$ can vary from 1 to $n^2$.

The algorithm leverages an oracle $\textsc{PatternMatchingOracle}$ that solves wildcard pattern matching in almost linear time and linear space \cite{Fischer1974STRINGMATCHINGAO, KalaiDontCares, CLIFFORD200753, Indyk98}. In particular, $\textsc{PatternMatchingOracle}$ returns a binary array which has size equals to the text size, with ones on all possible match locations. At a high level, our algorithm samples both the text and the pattern, and performs the oracle on the sampled strings. If the pattern matches the text, then the corresponding samples must also match. 
The sample operation on a string $t$ with parameters $k$ and \textit{offset} is defined as follows: 
\[ \textsc{Sample}(t, \textit{offset}, k) = t[\textit{offset}, \textit{offset}+k, \textit{offset} + 2k, \ldots ] \]
In other words, $\textsc{Sample}(t, \textit{offset}, k)$ is a string obtained by selecting all characters with indices $i > \textit{offset}$ in $t$ such that $i \mod k = \textit{offset} \mod k$. Note that $\textsc{Sample}(t, \textit{offset}, k)$  has size $O(|t|/k)$.
Therefore, by sampling the strings at an appropirate frequency $k$, the space usage can be bounded accordingly.

\begin{algorithm}[h!]
	\SetAlgoLined
	$k = \lfloor n/s \rfloor$; \\
	\For{\text{offset from 1 to $k$}} {
		\textit{res} = zero-array of size $n/k$; \\
	 	\For{\text{shift from 1 to $k$}}{
	 		\textit{sampled\_pattern} = \textsc{Sample}(\textit{pattern}, \textit{shift}, $k$); \\
	 		\textit{sampled\_text} = \textsc{Sample}(\textit{text}, \textit{offset + shift}, $k$);  \\ 
	 		\textit{res} = \textit{res} $\land$ \textsc{PatternMatchingOracle}(\textit{sampled\_pattern, sampled\_text});
	 	}
 		\If{res[i] = 1 for some i}{
 			\text{Return a match at position \textit{offset} + $ik$};
 		}
	} 
	\text{Report no match exists between} \textit{pattern} and \textit{text}; 
	\caption{Exact wildcard pattern matching in $\tilde O(n^2/s)$ time and $O(s)$ space.} \label{alg:non-streaming-pattern-matching}
\end{algorithm}

\begin{theorem} \label{thm:non-streaming-wildcard-pattern-matching}
	There exists an $\tilde O(n^2/s)$ algorithm for wildcard pattern matching with space $O(s)$ in the non-streaming setting. 
\end{theorem}

\begin{proof}
Suppose \textit{pattern} matches \textit{text} at position $l$. Let $$\textit{offset} = l \mod k$$ and $$i = ( l - \textit{offset} ) / k.$$
	The sampled pattern must also match the sampled text at position $i$ for all possible values of \textit{shift}. Therefore, if no match has been found, and Algorithm 2 proceeds to $\textit{offset} = l \mod k$, $\textit{res}[i]$ must be $1$ at this round. 
	Hence, it correctly outputs a match position if exists.
	
	Notice that each of the sampled strings has size at most $n/k$. Therefore, each call of the oracle takes $\tilde O(n/k)$ time and $O(n/k)$ space. The result of each call is combined with \textit{res}, which also has size $n/k$. Therefore, the total time is $\tilde O(n/k) \times k^2 = \tilde O(nk)$ and the required space is $O(n/k)$. Substituting $k = n/s$ gives the desired bounds.  
\end{proof}
\section{Longest Common Substring}
The main result of this section is an asymmetric streaming algorithm with memory $O(1)$ that runs in polynomial time. More precisely, our algorithm reads the input string in a single pass, runs in time $\tilde O(nm)$ where $n$ and $m$ are the lengths of the input strings and computes the solution with $O(1)$ memory. 

We begin by stating the main result of this section. In what follows, we give an asymmetric streaming algorithm for longest common substring that only uses memory $O(1)$ and solves the problem in polynomial time. The idea for this algorithm is a generalization of the technique we used for pattern matching. At each point in time, we keep two pointers to characters of $T$. We denote these two pointers by $\ell$ and $r$. These two pointers denote the largest substring of $T$ which is equal to a suffix of $S$. Note that $S$ is empty initially and as its characters arrive, its length increases over time. In addition to the above, we also keep an integer number which takes the maximum of $r - \ell +1$ throughout the lifetime of the algorithm. That is, we keep the maximum solution that we find over all steps of our algorithm.

Once a new element arrives at the end of $S$, we need to update the two pointers $\ell$ and $r$. Although we only spend memory $O(1)$, we have access to all the necessary information about $S$. Two pointers $\ell$ and $r$ in addition to the newly arrived element give us a postfix of $S$ and due to the guarantee of our algorithm, the new postfix of $S$ which we need to find certainly lies inside this substring. Thus, all the necessary information is available with $O(1)$ memory and the only task is to update pointers $\ell$ and $r$. Before we state Theorem~\ref{thm:asym-stream-longest-common-substring} we would like to emphasize that $\ell$ and $r$ may not represent the longest common substring between $S$ and $T$ at every point in time. That is, it is quite possible that at some point, the interval $T[\ell, r]$ gives the longest substring of $T$ which is a postfix of $S$ but the longest common substring of $S$ and $T$ is larger. This happens when the solution is not necessarily a postfix $S$. However, since we iteratively maintain such a solution for each prefix of $S$, at some point in time the optimal solution to longest common substring is also a postfix of $S$ in which case $T[\ell,r]$ is in fact the longest common substring of $S$ and $T$. Since we report the maximum size of all such solutions, our algorithm always reports the size of the longest common substring. We bring a formal proof for this algorithm in Theorem~\ref{thm:asym-stream-longest-common-substring}.

\begin{algorithm}[h!]
	\SetAlgoLined
	\KwResult{}
	mx $\leftarrow 0$\;
	$\ell \leftarrow -1$\;
	$r \leftarrow -1$\;	
	\While{new character comes}{
		$\ell,r \leftarrow $ new $\ell$ and $r$ computed by the pattern matching algorithm\;
		mx $\leftarrow \max\{$output$, r - \ell+1\}$\;
	}
	Report mx\;
	\caption{Exact asymmetric streaming algorithm for \textsf{LCS}}\label{alg:lcs1}
\end{algorithm}

\begin{theorem} \label{thm:asym-stream-longest-common-substring}
	There exists a randomized algorithm for asymmetric streaming \textsf{LCS} that solves the problem with memory $O(1)$ and runtime $\tilde O(nm)$ where $n$ and $m$ denote the size of the two strings, respectively. This algorithm succeeds with probability at least $1-n^{-5}$.
\end{theorem} 
\begin{proof}
As explained earlier, our algorithm only keeps track of 3 variables $\ell$, $r$, and $mx$ where $mx$ keeps the maximum of $r-\ell+1$ over all steps of the algorithm. initially, $S$ is empty so we set both $\ell$ and $r$ to $-1$ meaning that our initial solution is empty. At every point in time $T[\ell, r]$ determines the largest postfix of $S$ which also appears in $T$ as a substring. The key to our algorithm is the following: Let $S'$ be $S$ after adding a new character. If $Q$ and $Q'$ are the largest postfixes of $S$ and $S'$ that appear in $T$ as substrings, then we have $|Q'| \leq |Q|+1$. To prove this, we assume for the sake of contradiction that $|Q'| > |Q|+1$. This means that aside from the newly added character to $S$, the new postfix of $S'$ that is shared with $T$ is larger than $|Q|$. This obviously contradicts with the maximality of $Q$.
	
	The above observation shows that the only relevant parts of $S$ that we need to have access to in order to update the solution are the last $r-\ell+2$ characters. Moreover, $T[\ell,r]$ gives us $r-\ell+1$ characters of this list and therefore by keeping the newly added character with memory $O(1)$ we can have full access to all such characters. 
	
	The next task is to update $\ell$ and $r$. In other words, we need to find the largest postfix of $S$ that appears as a substring in $T$. Based on the above observation, we define $Q = T[\ell,r]$ and $Q^{+}$ which is equal to $Q$ plus the newly added element appended to the end of $Q$. As explained earlier, access to every element of $Q^{+}$ is available in $O(1)$ time. In order to find the largest postfix of $Q^{+}$ that appears in $T$ as a substring, we use our asymmetric streaming algorithm for pattern matching in the reverse manner. In other words, we begin by $Q' = \emptyset$ initially, and add the elements of $Q^{+}$ to $Q'$ from right to left. In other words, in the first step $Q'$ only contains the last element of $Q^{+}$. Next, we add the second last element of $Q^{+}$ and continue on. Recall that our asymmetric streaming algorithm for pattern matching requires memory $O(1)$ and linear time and therefore with the same complexity we can find the largest postfix of $Q^{+}$ that appears in $T$ as a substring. Therefore, the total runtime for updating the solution is $O(|S|+|T|)$.
	
Since we run the above procedure every time a new character is added to $S$, then the overall runtime is bounded by $O(|S|^2 |S||T|)$.
\end{proof}

Algorithm~\ref{alg:lcs1} gives a perfect solution in terms of memory and accuracy. Although the algorithm takes quadratic time to solve \textsf{LCS}, it seems that there is an inherent difficulty in obtaining a subquadratic time solution for \textsf{LCS} in the asymmetric streaming model. To see this, consider the following special case of \textsf{LCS}: To strings are given as input, and the goal is to figure out if the two strings have any character in common? Of course, a solution for \textsf{LCS} immediately gives a solution for the above problem since the solution of \textsf{LCS} is greater than 0 if and only if the two strings have at least a character in common.

Although no technique is known to prove unconditional lower bounds on the time complexity of problems, it seems that with additional memory $O(1)$, no algorithm can do better than searching the entire offline string once a new character comes in the online string. This indeed takes time $O(n^2)$ which is equal to the runtime of our algorithm for \textsf{LCS}. This signals that the only way to improve the runtime of the algorithm is to make an assumption on the solution size to give more flexibility to our algorithm for cases where the solution size is at least $\omega(1)$ (the above argument works only if the solution size is $1$).

We complement the above observation with an approximation algorithm that \textsf{LCS} with constant memory and sublinear update time for large solutions. We show that for any $\epsilon, \kappa > 0$, there exists an algorithm for \textsf{LCS} that approximate the solution within a factor $1-\epsilon$ in $O(1/\kappa)$ rounds and its overall runtime is bounded by $\tilde O(n^{2+\kappa}/d)$. Notice we can set $\epsilon$ and $\kappa$ arbitrarily close to $0$ while keeping them constant. This results in an asymmetric streaming algorithm that runs in constant rounds.

\begin{theorem}\label{theorem:lcs2}
	For any $\epsilon, \kappa > 0$, there exists an algorithm for \textsf{LCS} that approximate the solution within a factor $1-\epsilon$ in $O(1/\kappa)$ rounds with memory $O(\log n/\epsilon)$ and its overall runtime is bounded by $\tilde O(n^{2+\kappa}/d)$ in the worst case.
\end{theorem}
\begin{proof}
We start by making an extra assumption that simplifies our solution. At the end we discuss how to modify our algorithm to make it independent of our assumption. We assume for simplicity that the value $d$ is give to us as input and we just need distinguish the following two cases:
\begin{itemize}
	\item The solution size is at least $d$.
	\item The solution size is bounded by $(1-\epsilon)d$.
\end{itemize} 
	
With this extra assumption, we design our algorithm in the following way: We mark $n/(\epsilon d)$ characters of $T$ that evenly divide the string. In other words, each pair of consecutive marked characters of $T$ are $(\epsilon d)$-away. Moreover, in our solution, we only taking into account common subsequences of the two strings such that the position of it's first character in $T$ is marked. With this assumption we only lose a multiplicative factor $1-\epsilon$ in the accuracy of the algorithm since if the \textsf{LCS} size is at least $d$, one can ignore a prefix of size at most $\epsilon d$ of the solution such that the remainder starts from a marked character of $T$.

To solve \textsf{LCS}, every time a marked character arrives, we run a pattern matching algorithm  that finds a pattern starting from the marked character with length $(1-\epsilon) d$ in $S$. For this purpose we run the algorithm of Theorem~\ref{thm:asym-stream-pattern-matching} that uses only $O(1)$ memory and runs in time $\tilde O(n)$. Since the length of each pattern is $(1-\epsilon d)$ and the marked characters are $\epsilon d$ characters away, then at each point in time we simultaneously solve $O(1/\epsilon)$ pattern matching problems and therefore the memory of our algorithm is $O(1/\epsilon)$. Also, since the runtime of the pattern matching algorithm is $\tilde O(n)$, then the overall runtime of our algorithm is $\tilde O(n^2/(\epsilon d)$ since we solve $O(n/(\epsilon d))$ instances of the pattern matching problem.

The typical way to make our algorithm oblivious to the solution size is to start with assuming $d = n$, and each time try to find a solution of size $(1-\epsilon) d$ with the above algorithm. If we fail to do so, we multiply $d$ by a factor $1-\epsilon$ and repeat our algorithm with the new guess for the solution size. While this comes with no overhead on the memory of the algorithm and a multiplicative overhead of at most $O(\log n/\epsilon)$ for the runtime, the algorithm requires a logarithmic number of passes over the input.

Another typical approach is to apply the above trick but run all instances simultaneously. This way, the number of passes for the algorithm remains 1, but the memory of the algorithm is multiplied by an extra $O(\log n / \epsilon)$ factor. The more serious issue with this approach is that the runtime of the algorithm may no longer be $\tilde O(n^2 / d)$ due to the following observation: While for the correct guess of $d$, our algorithm only needs time $\tilde O(n^2/(\epsilon d))$ time to find the solution, it may require more time for smaller values of $d$. In particular, for smaller guesses ($d = O(1)$) our algorithm requires quadratic time and since we solve all the instances simultaneously our algorithm is also running those cases. Thus, it is quite possible that while there is a solution of size $d$ in the second half ef $T$, our algorithm requires a runtime of $\tilde O(n^2)$ for processing the first half of $T$ and thus before any characters of the solution arrive, we need to spend time $\tilde O(n^2)$.

To resolve the above issue, we take a hybrid approach: We divide the $O(\log n/\epsilon)$ guesses for $d$ into $1/\kappa$ categories where in each category the values of $d$ differ by at most a multiplicative factor $n^{\kappa}$. Thus, in each category the number of problem instances is $(\kappa \log n)/ \epsilon$. We start from the category with the highest guesses and each time run all the instances of that category simultaneously. If we find a solution for any value of $d$, we report the largest solution found. Otherwise, we proceed to the next highest category.

With the new approach, the runtime for a category with highest guess $d$ is equal to $\tilde O(\kappa n^{2+\epsilon}/(\epsilon d))$. Moreover, the memory of each run is bounded by $O(\kappa \log n/\epsilon)$. Since there are $O(1/\kappa)$ categories, the overall runtime is $\tilde O(n^{2+\epsilon}/(\epsilon d))$ and our algorithm makes at most $O(1/\kappa)$ passes over the input. Also, the memory of our algorithm is $O(\kappa \log n/\epsilon)$	
\end{proof}

By setting $\kappa = 1-1/\log n$, Theorem~\ref{theorem:lcs2} turns into an algorithm with runtime $\tilde O(n^2/d)$ that runs in $O(\log n/\epsilon)$ rounds.

\begin{corollary}[of Theorem~\ref{theorem:lcs2}]\label{cor:lcs3}
	For any $\epsilon> 0$, there exists an asymmetric streaming algorithm for \textsf{LCS} that approximate the solution of \textsf{LCS} within a factor $1-\epsilon$ in $O(\log n/\epsilon)$ rounds with memory $O(1/\epsilon)$  and its overall runtime is bounded by $\tilde O(n^2/d)$ in the worst case.
\end{corollary}

\bibliographystyle{apalike} 
\bibliography{main}

\end{document}